\documentclass[aps, 12pt, amsmath, amssymb, noshowpacs, nofootinbib, preprint, preprintnumbers, prd]{revtex4-1}
\usepackage{amsmath}
\usepackage{amsfonts}
\usepackage{amssymb}
\usepackage{textcomp}
\usepackage[dvips]{graphicx}
\usepackage{graphicx}
\usepackage{bm}
\usepackage{color}

\usepackage{epsfig}

\usepackage{epsfig}
\usepackage{graphicx}

\usepackage{latexsym}
\usepackage{amsmath}
\usepackage{amssymb}
\usepackage{relsize}
\usepackage{geometry}
\usepackage{amsfonts,amsmath,amssymb}

\newcommand{\be}{\begin{equation}}
\newcommand{\ee}{\end{equation}}
\newcommand{\ba}{\begin{eqnarray}}
\newcommand{\ea}{\end{eqnarray}}
\newcommand{\ben}{\begin{enumerate}}
\newcommand{\een}{\end{enumerate}}

\newcommand{\gm}{\gamma_{\mu}}
\newcommand{\gn}{\gamma_{\nu}}
\newcommand{\dmu}{\partial_{\mu}}

\newcommand{\dl}{\partial_{\lambda}}

\newcommand{\dmuu}{\partial^{\mu}}

\newcommand{\dz}{\partial_{z}}

\newcommand{\dt}{\partial_{0}}
\newcommand{\dzed}{\partial_{3}}
\newcommand{\lb}{\left(}
\newcommand{\rb}{\right)}
\newcommand{\lbf}{\left\{}
\newcommand{\rbf}{\right\}}
\newcommand{\ld}{\left.}
\newcommand{\rd}{\right.}
\newcommand{\lv}{\left|}
\newcommand{\rv}{\right|}
\newcommand{\lbr}{\left[}
\newcommand{\rbr}{\right]}
\newcommand{\ltr}{\left\langle}
\newcommand{\rtr}{\right\rangle}

\def\beq{\begin{equation}}
\def\eeq{\end{equation}}
\def\beqn{\begin{eqnarray}}
\def\eeqn{\end{eqnarray}}

\newcommand{\gsim}{\lower.7ex\hbox{$
\;\stackrel{\textstyle>}{\sim}\;$}}
\newcommand{\lsim}{\lower.7ex\hbox{$
\;\stackrel{\textstyle<}{\sim}\;$}}

\def\slashed#1{\setbox0=\hbox{$#1$}             
   \dimen0=\wd0                                 
   \setbox1=\hbox{/} \dimen1=\wd1               
   \ifdim\dimen0>\dimen1                        
      \rlap{\hbox to \dimen0{\hfil/\hfil}}      
      #1                                        
   \else                                        
      \rlap{\hbox to \dimen1{\hfil$#1$\hfil}}   
      /                                         
   \fi}                                        %

\begin{document}

\begin{titlepage}
\thispagestyle{empty}

\begin{flushright}
FTPI-MINN-12/01, UMN-TH-3027/12\\
ITEP-TH- 41/11
\end{flushright}

\vspace{1cm}

\begin{center}
{  \Large \bf More on the Tensor Response of the QCD Vacuum to \\
[2mm] an External Magnetic Field }
\end{center}
\vspace{1mm}

\begin{center}

 {\large
 \bf   A.\,Gorsky,$^{\,a}$  P.\,N.\,Kopnin,$^{\,a,c}$ A.\,Krikun,$^{\,a}$  and \bf A.\,Vainshtein$^{\,b}$}

\vspace{3mm}

$^a${\it Theory Department, ITEP, Moscow, Russia}\\[1mm]
$^b${\it  William I. Fine Theoretical Physics Institute,\\
University of Minnesota,
Minneapolis, MN 55455, USA}\\[1mm]
$^c$
{\it Moscow Institute of Physics and Technology, Dolgoprudny, Russia}
\end{center}

\vspace{1cm}

\begin{center}
{\large\bf Abstract}
\end{center}
 In this Letter we discuss a few issues concerning
the magnetic susceptibility of the quark condensate
and the Son-Yamamoto (SY) anomaly matching equation. It is shown
that the SY relation in the IR implies a nontrivial interplay
between the kinetic and WZW terms in the chiral Lagrangian.
It is also demonstrated that in a holographic framework an external magnetic
field triggers mixing between scalar and tensor fields. Accounting for this,
one may calculate the magnetic susceptibility of the quark condensate to
all orders in the magnetic field.

\end{titlepage}

\newpage


\section{Introduction}
The magnetic susceptibility $\chi$, introduced in \cite{is}
in the sum rules framework, is
an interesting characteristic of the vacuum response to
an external magnetic field in the confinement phase. It measures
the induced tensor current in the QCD vacuum. The expression
\begin{equation}
\label{chi_v}
\chi_v = - \frac{N_c}{4 \pi^2 f_{\pi}^2}
\end{equation}
has been obtained
analytically in \cite{vai} using the
OPE and pion dominance for the $\langle VVA\rangle$ correlator of two
vector currents and one axial current in the kinematics where two
virtualities of the external legs are large while one
vector current represents the constant external field strength. Surprisingly
it differs from the sum rule fit by a factor of 3 which
implies that some qualitative essential effect responsible for the
disagreement has been overlooked yet. The several
phenomenological estimates yield the low value of the
susceptibility  while the large $N_c$ consideration \cite{cata09}
fits the Vainshtein relation (\ref{chi_v}).

Hence it is natural to look for alternative derivations
of $\chi$ to identify the missed ingredients. The problem
was discussed in the holographic hard wall model involving
the 5D Yang-Mills and Chern-Simons (CS) terms. The Lagrangian
corresponds to the gauge theory on the flavor branes extended
along the radial coordinate in the AdS space. It turns out
that in this model the Vainshtein relation (\ref{chi_v}) is not exact,
however it is fulfilled with good accuracy \cite{gk}.
Moreover it was shown in \cite{gk} that
the whole answer follows from the CS term which
implies that we are dealing with a sort of ``anomalous" phenomena.

Other more refined holographic models have been considered by Son and
Yamamoto (SY) in \cite{sy}. They derived a new relation between two-point
and three-point correlators  which yields nontrivial matching
conditions for the low-energy QCD parameters of the mesons.
The SY relation is assumed to be valid at any momentum transfer, for instance, Vainstein expression
(\ref{chi_v}) follows from the SY relation at large virtualities if one assumes that operator product
expansion of QCD is applicable.
This `if' is important because the SY model does not support OPE {\em per se}:
dependence on momentum transfer is exponential and does not contain power
terms required by OPE.
 Different aspects of the SY relation were discussed
in \cite{deraf,cola,kiritsis}.

The situation looks a little bit puzzling since there is no
field theory derivation of the SY relation yet. The expression
for $\chi$ in terms of pion decay coupling suggests that
it can be obtained purely in terms of the chiral Lagrangian
together with the SY relation.  With the holographic
experience it could be expected that the Wess-Zumino-Witten (WZW)
term in the chiral Lagrangian related to the CS term in 5D should be responsible 
for the nontrivial answer.

In this note motivated by the comments above we consider
the additional arguments concerning the derivation of $\chi$.
Since the holographic model of QCD is nothing but the extended
chiral Lagrangian it is natural to look more carefully at the
place of the SY relation in the ChPT per se. The small $Q^2$
region is the most comfortable to be  analyzed in ChPT hence we shall
look at the first terms in   small $Q^2$ expansion. The relation between
the ChPT parameters at the tree level has been discussed in \cite{deraf}
and we extend it to the one loop level focusing at the chiral logs.
It will be shown that the SY relation holds true for the log terms.

A simple argument involving the calculation of the quark determinant
in the tensor source background  implies that the nonvanishing magnetic
susceptibility corresponds to the peculiar additional mixed term in the Lagrangian.
In the improved holographic model for QCD \cite{cata,harvey,Karch:2011} the tensor
source in $D=4$ theory is promoted into the tensor field in $D=5$. We
shall analyze the improved model in a  magnetic field focusing
at the scalar-tensor mixing. It turned out that in the improved model
the magnetic susceptibility can be obtained to all orders in the
magnetic field. In a small field the magnetization grows linearly with the field, in accordance with its generally established properties, while in a large field it does not depend on the magnetic field.

The paper is organized as follows. In Section 2 we consider
the SY relation in ChPT and show  the matching of the
chiral logs in this relation. Some general comments
concerning the SY relation are also presented.
In Section 3 we consider the improved holographic model
for QCD with the tensor field   and demonstrate how the scalar-tensor mixing
in the magnetic field  yields the magnetic susceptibility.
Some discussion concerning the proper degrees of freedom
can be found in the last Section.

\section{SY relation within the Chiral Lagrangian}
\subsection{SY relation and chiral logs}
Let us analyze the
SY relations in the framework of ChPT. The SY relation
has been obtained holographically and is based on
the simple $D=5$ action on the worldvolume of  probe
flavor branes involving Yang-Mills and
Chern-Simons terms for the flavor gauge group. Taking into
account that correlators of the vector and axial currents
are two independent solutions to the second order differential
operator in the radial coordinate $z$  in AdS space their
Wronskian is $z$-independent. This argument works when the CS
term is neglected. On the other hand CS term itself yields the nontrivial
$\langle VA \rangle$ correlator in the magnetic field which is proportional
to the same Wronskian. Hence in a weak magnetic field the following
relation holds \cite{sy}
\beq
w_{T}(Q^2)= \frac{N_C}{Q^2} -
\frac{N_C}{f_{\pi}^2}\,\Big[\Pi_A(Q^2) - \Pi_V(Q^2)\Big]
\label{sy}
\eeq
where $w_{T}$ is defined via the two-point correlator
in the external weak electromagnetic field with the constant field strength $F_{\alpha\beta}$
\beq
\langle V_{\mu}A_{\nu} \rangle_{\tilde{F}} = \frac{1}{4\pi^2}
\big[w_T(q^2)(-q^2 \tilde{F}_{\mu\nu} + q_{\nu}q^{\sigma} \tilde{F}_{\mu\sigma} -
q_{\mu}q^{\sigma} \tilde{F}_{\nu\sigma}) + w_L (q^{2}) \,q_{\nu}q^{\sigma} \tilde{F}_{\mu\sigma}\big],
\eeq
${V,A}$ are the  vector, $\bar q\,{\cal V}\gamma_\mu q$, and axial, $\bar q\,{\cal A}\gamma_\mu \gamma_5 q$, currents, $\tilde F$ denotes the dual field strength, $\tilde F_{\gamma\delta}=\slantfrac{1}{2}\,\epsilon_{\gamma\delta\alpha\beta}F^{\alpha\beta}$, and
$\Pi_A\,,\,\Pi_V$ are the corresponding two-point correlators.
The relation holds
for all values of $Q^2=-q^{2}$.

The definitions of the above correlators are
\begin{eqnarray}
\frac{1}{2}\,{\rm Tr}\,({\cal QVA})\,\langle V_{\mu}A_{\nu} \rangle_{\tilde{F}}\equiv
\int d^4x \,e^{iqx}\langle T \{V_\mu(x) A_\nu(0)\}\rangle_{\tilde{F}}\,,\nonumber\\
\frac{1}{2}\,{\rm Tr}\,({\cal VV}) \,\Pi_V(Q^2)(q_\mu q_\nu-g_{\mu\nu} q^2)\equiv
\int d^4x \, e^{iqx}\langle T \{V_\mu(x) V_\nu(0)\}\rangle_{0}\,,\\
\frac{1}{2}\,{\rm Tr}\,({\cal AA}) \,\Pi_A(Q^2)(q_\mu q_\nu-g_{\mu\nu} q^2)\equiv
\int d^4x \, e^{iqx}\langle T \{A_\mu(x) A_\nu(0)\}\rangle_{0}\,, \nonumber
\end{eqnarray}
where flavor dependence on matrices of vector and axial currents, ${\cal V}$ and ${\cal A}$,
as well as  that for the electric charge, $\cal Q$, is factored out.

It is natural to look at the matching of the SY relation with the ChPT
since the holography provides the ChPT derivation from the ``first principles".
The SY relation is derived from the $D=5$ equations of motion hence according
the holographic dictionary it should correspond to the tree approximation
in the ChPT. The condition imposed by the SY at $Q^2=0$ on the parameters
of the chiral lagrangian has been found in \cite{deraf}
\beq
L_{10}=- 4\pi^2 C_{22}
\label{tree}
\eeq
where $L_{10}$ corresponds to  the even term in the chiral Lagrangian
at the $O(p^4)$ order while $C_{22}$ corresponds to a particular
odd term at the $O(p^6)$ order. The condition is unexpected since it
relates the odd and even terms in the Lagrangian. Unfortunately
this relation between constants can not be used as the test
of the SY relation since $C_{22}$ is not known with the high accuracy.

To get some test of the SY relation let us focus at the terms in the
correlators involving the chiral logs. Some comments are required
before the looking at log terms. Naively such terms are subleading in $1/N_{c}$
hence in the holographic approach these should be considered as
corrections to the equations of motion.  On the other hand the
log terms are considered as the renormalization of the constants
in the chiral Lagrangian hence one should assume that the
relation (\ref{tree}) valid at the tree level holds upon the
renormalization. Therefore it is natural to look at the matching
of the chiral logs.

At the right hand side of Eq.\,(\ref{sy})  the chiral log follows from
the pion loop in the correlator of the vector currents
\beq
\Pi_V^{\rm chir}(Q^2\rightarrow 0)= c \log Q^2\,, \qquad c=-\frac{1}{48\pi^2}\,.
\eeq
There are no logs in the correlator of the axial currents.
On the other hand the chiral log in the $\langle VVA \rangle$ correlator
can be traced  from the particular term in the WZW term in the chiral Lagrangian
responsible for the decay $\gamma^{*} \rightarrow 3\pi$,
\begin{equation}
S_{WZW}^{3\pi}=-\frac{N_c}{24\pi^2}\int {\rm Tr}\, A\big(dU^{-1}U\big)^3\to
-\frac{iN_c}{24\pi^2 f_\pi^3}\int {\rm d}^4 x\,{\rm Tr}\, \tilde F^{\gamma\delta}\pi
\,\partial_\gamma \pi \,\partial_\delta \pi \,.
\end{equation}
Converting two pions from this vertex to the vector current and associating the remaining pion
with the axial current we get the $\log Q^2$ contribution to $w_T$\,,
\beq
w_T^{\rm chir} =c_1 \log Q^2\,, \qquad
c_1= \frac{N_c}{f_{\pi}^2}\,c\,,
\eeq
which is consistent with the SY relation.

Let us emphasize that there is no freedom in the terms
in the chiral lagrangian involved into the chiral logs, hence
the matching is exact although at the subleading order in $1/N_c$.
Note that similarly to the tree-level case, the
SY relation  implies
an unexpected relation  between the coefficient in front of the
even kinetic  term at the $O(p^2)$ order and of an odd WZW term at the
$O(p^4)$ order in the chiral Lagrangian.

\subsection{The mixed  term in the chiral Lagrangian}
Let us argue  that nonvanishing  magnetic susceptibility implies
a peculiar  term in the effective Lagrangian.
To this  aim we introduce a source
term for the quark tensor current into the QCD Lagrangian
\beq
\label{bmunu}
\delta L_1 = B_{\mu\nu}\,\bar q \,{\cal B}\,\sigma^{\mu \nu} q\equiv 
 i\tilde B_{\mu\nu}\,\bar q \,{\cal B}\sigma^{\mu \nu}\gamma_{5} q
\eeq
where $B_{\mu \nu}$ is an external source field whose possible interpretation
shall be discussed below and ${\cal B}$ is a diagonal flavor matrix. 
Accounting for the chiral features of the quark operator in Eq.\,\eqref{bmunu}
it simple to determine the corresponding term in the chiral Lagrangian
in the linear approximation in the $B_{\mu \nu}$ field,
\beq
\delta L_{WZW} = -\frac{1}{2}\,\chi \,\langle\bar{q} q\rangle\,B_{\mu \nu} F^{\mu \nu}\, {\rm Tr}\, (U+U^{\dagger}){\cal B}{\cal Q} \,,\label{WZWterm}
\eeq
where $\chi$ is the magnetic susceptibility, $\langle\bar{q} q\rangle$ is the quark condensate and $U=\exp (2i\pi^a t^a/f_{\pi})$ is the mesonic matrix 
($f_{\pi}=92 \,{\rm MeV}$). This can be viewed as a definition of the magnetic susceptibility.
Note that this term to some extent can be considered
as the shift of the effective quark mass in the external fields. In the
next Section we shall see that this term promoted into the
holographic $D=5$ action provides an important scalar-tensor
mixing.

It it worth making a few comments concerning the implications of  this effective WZW-like
term in the chiral Lagrangian.  First, the vacuum tensor current proportional
to the chiral condensate in the magnetic field can be attributed to the
stringy degrees of freedom if we identify the tensor source in (\ref{WZWterm}) as
NS or Ramond two-form fields. With such identification the flow of the
F1 or D1 strings in the vacuum occurs in the magnetic field. On the other
hands the mesons are identified holographically as the F1 strings
connecting the flavor branes  hence such ``stringy" current corresponds
in fact to a kind of mesonic vacuum current.

Secondly, there  is an anomalous electromagnetic
current proportional to the condensate
in the external tensor field. Indeed we defined the current
\beq
\langle J_{\nu}\rangle_{B} =\frac{\delta S_{WZW}}{\delta A_{\nu}}
\eeq
which gets contributions from the sources of the tensor field or
from the varying pion field.
\beq
\ltr J_{\nu} \rtr_{B} =\frac{1}{2}\,\chi \,\langle\bar{q} q\rangle\,  \partial^{\mu}\big[B_{\mu\nu} {\rm Tr}\, (U+U^{\dagger}){\cal B}{\cal Q}\big]. \label{current_chipt}
\eeq
For the varying pion case  the anomalous current
is the analogue of the Goldstone-Wilczek current. For the varying tensor field
an interesting possibility emerges. Using the relation between the massive vector
and tensor in four dimensions we could get the nonvanishing electromagnetic
current if the vector meson gets condensed. There are some indications of such a
condensation in the magnetic field both in the effective theory \cite{chernodub}
and in the holographic framework \cite{erd}. Hence one could speculate about
the nonperturbative current proportional to the product of quark and  vector meson condensates.

Finally, if we expand
the anomalous term in the pion field we could get the anomalous interaction
of pions with the tensor current in the magnetic field. For instance, the matrix element
\beq
\langle 0|\bar{q}\sigma_{\mu \nu} q|\pi^{0}\pi^{0}\rangle =\frac{1}{3f_{\pi}^{2}}\,\chi\, \langle \bar{q}q\rangle
 F_{\mu \nu}\,.
\eeq

\subsection{On the derivation of the SY relation}
The SY relation has been obtained in a slightly tricky way, hence
it would be nice to get it  more regularly as a kind of a
Ward identity. Here we restrict ourselves by two generic remarks.
Since the key observation in the derivation in \cite{sy} was the
$z$-invariance of the Wronskian of the vector and axial currents
it is reasonable to look at the radial  variable $z$ in the Hamiltonian
framework. That is, following \cite{vvdb}  we assume that it is considered
as a time variable for the
RG Hamiltonian evolution in the $D=5$ gauge theory.

In the Hamiltonian framework of the gauge theories there are
two natural equations involving the dependence on the boundary
values of the dynamical variables. These are the gauge constraint
or Gauss law and the Hamiltonian constraint or a kind of the
Hamilton-Jacobi (HJ) equation. We are in a peculiar situation
with the Hamiltonian constraint since the metric  depends
on the radial coordinate and is therefore ``time-dependent".

First consider the Gauss law constraint with respect to the
flavor gauge group ${\rm SU}_L(N_F)\times {\rm SU}_R(N_F)$ on the flavor
branes. In the Hamiltonian approach the Gauss law reflect
the gauge invariance  with respect to the flavor gauge group
and can be identified with the generator of the z-independent gauge
transformations.
Since there are $D=5$ CS terms
for the left and right gauge flavor fields the canonical momenta get modified
\beq
\Pi_L= E_L + A_LF_L \,,\qquad              \Pi_R= E_R - A_LF_L\,,
\eeq
while the canonical momenta for the scalar field are standard. Using
the Hamiltonian relation
\beq
\Pi =\frac{\delta S}{\delta x}\,,
\eeq
where S is the action, and the standard holographic relation for the $D=4$ currents
$J_{\mu}=\frac{\delta S}{\delta A}$  one
immediately recognizes that the Gauss law constraint in the
bulk theory precisely produces the anomaly equation
for the axial current at the boundary
including the mass term. The fact that the Gauss law is valid at any
time in $D=5$ theory  gets translated into the claim that the axial anomaly is seen at all scales in
the boundary $D=4$ theory. Note that in the conventional gauge
theory the Gauss law is complemented by the gauge $A_0=0$. In the
current situation the similar equation reads as $A_z=0$; however,
one should not forget that the pion field can be identified
with the holonomy of the radial component of the flavor gauge group.

In the holographic setting the
HJ equation for the bulk metric  has been identified as the RG equation
in the boundary theory in \cite{vvdb}. Here we have to consider
a similar HJ equation for the gauge fields and scalars. Taking into account
the shift of the canonical momenta and forgetting for a moment
the metric  one obtains for the left gauge part of the total Hamiltonian
\beq
\Big(\frac{\delta S}{\delta A} - AF\Big)^2 + F_{ij}^2
\eeq
and similarly for the contribution of the right gauge field and scalars.
The HJ-type equations are quite convenient for the derivation of
Ward identities in the boundary theory since it involves the desired
variational derivaties.
It is important that the HJ-like equations due to the change
of the canonical momenta
involve the  terms with the different number of the
variational derivatives. Hence potentially one could hope
that the additional variational derivatives of the HJ equation
upon taking into account the Gauss law constraint would yield
the SY relation. We did not succeed along this way of
reasoning, however we plan to discuss the complete set of the
Ward identities induced from the bulk theory elsewhere.
In particular we plan to elaborate the constraints emerged
from the dynamics of the higher rank fields induced by the
color branes in the brane approach.
Note that some examples of the derivation of the
boundary Ward identities from the bulk HJ equation 
can be found in \cite{corley}.

There is also some  analogy with the $N=1/2$ SUSY YM case which
can be considered as $N=1$ SYM theory in the self-dual constant
graviphoton background $C_1$. The following term gets induced in the
graviphoton field
\beq
\delta L = dC_1\wedge F \bar{\lambda \lambda}
\eeq
which is analogous to the anomalous term in QCD we have discussed.
The analogy with QCD becomes even more close when we remind that
the gluino condensate is developed in $N=1/2$ SUSY YM like
the chiral condensate in QCD.  Moreover
in the $N=1/2$ theory one can consider the Ward identities
reflecting the single unbroken SUSY \cite{iman}. This
Ward identity amounts to a particular degeneration in the
spectral densities in the $J=1^{\pm}$ channels \cite{gs}.
Since the spectral densities follow from the two-point correlators
these Ward identities can be considered as some analogue
of the SY relation in QCD without the anomalous three-point correlator.

\section{A Holographic Model with the tensor field}

In this Section we shall consider the scalar-tensor mixing in the
improved holographic model of QCD which involves the tensor field  \cite{cata,harvey,Karch:2011}.
It is a 5-dimensional gauge theory embedded in a pure AdS geometry with an infrared hard-wall boundary:

\be
ds^2 = \frac{\ell^2}{z^2}\lb - dz^2 + \eta_{\mu\nu}dx^{\mu}dx^{\nu} \rb,~~~\ 0 \leq z \leq z_m\,,
\ee
where $\eta_{\mu\nu}$ is mostly negative: $\eta = \rm diag(+---)$, and $\ell$ is the $AdS_5$ radius and shall be omitted hereforward (thus rescaling the coupling constants).
This model contains three types of fields: a complex scalar $X$, two gauge fields $L_{\mu}$ and $R_{\mu}$, and a complex antisymmetric tensor $B_{\mu\nu}$.
They are put into correspondence with the following operators of QCD:
\begin{align}
&\bar{q}_{R\,\bar f} \,q_L^{f} \leftrightarrow X_{\bar f}^f\,,
&\bar{q}_{R\,\bar g} \gm q_R^{\bar f} \leftrightarrow R_{\mu \,\bar g}^{\bar f}\,,\nonumber\\
&\bar{q}_{R\,\bar f}\,\sigma_{\mu\nu} q_L^{f} \leftrightarrow B_{\mu\nu\,\bar f}^{f}\,,
&\bar{q}_{L\,g} \gm q_L^{f} \leftrightarrow L_{\mu\, g}^{f}\,,
\end{align}
where $\sigma_{\mu\nu} = \frac{i}{2}[\gm,\gn]$; $f,\bar f$ are the flavor indices of QCD with respect to the (global) $U(N_f)_L\times U(N_f)_R$ symmetry which becomes the gauge group of the five-dimensional theory.
Accordingly, the fields $X$ and $B_{\mu\nu}$ are bifundamentals, whereas $L_{\mu}$ and $R_{\mu}$ are adjoint with respect to $U(N_f)_L$ and $U(N_f)_R$. These properties allow us to properly
introduce covariant derivatives: $DX = dX - i LX + iXR$, $H = DB = dB - i L\wedge B + i B\wedge R$.

The action proposed in \cite{Karch:2011} is:
\ba
{\cal S} &=& \int\ d^5x \sqrt{-g}\ {\rm Tr}\lbf -\frac{1}{4g_5^2}\lb F_L^2 + F_R^2\rb + g_X^2\lb |DX|^2 - m_X^2 |X|^2 \rb \rd \nonumber\\ [1mm] &-&\ld 2g_B\Big( \frac{i}{6}\lb B\wedge H^+
- B^+\wedge H \rb + m_B |B|^2 \Big) +\frac{\lambda}{2}\big( X^+F_L B + B F_R X^+ + {\rm c.c.} \big) \rbf.\label{fullaction}
\ea
This action is a modification of a simpler hard-wall action \cite{Erlich_Katz:2005} which takes into account the tensor field. The interaction term $XFB$ on the AdS boundary is reduced to the
term (\ref{WZWterm}) of the chiral Lagrangian which we have discussed in the previous Section.

The constants have been fixed in previous works by comparing
various correlators at large Euclidean $Q^2$ with OPE in QCD \cite{Erlich_Katz:2005,Cherman:2009,Krikun:2008,Karch:2011}.
The masses are fixed by requiring that the scaling properties of the fields match those of the corresponding operators in the UV: $m_X^2 = -3$, $m_B = 2$. Note that due to a non-canonical form
of the kinetic term of the tensor field its physical mass is actually $1$ in units of $\ell^{-1}$. In this case the vacuum solution for $X$ is
\be X(z) = \frac{1}{2}\Big( m z + \frac{1}{g_X^2} \ltr\bar{q}q\rtr z^3 \Big) \times \textbf{1}_{N_f\times N_f}\,.\label{Xmatching}\ee

From now on we shall only consider the Abelian degrees of freedom, as the flavor structure of the 5D fields is trivial, since the condensates, both scalar and tensor, as well as the electromagentic field, are diagonal in the flavor space. Hence, the equations of motion for each individual flavor $q_f$ are the same as for the singlet component with a substitution $F_{L,R}^{MN} \rightarrow e_f F_{L,R}^{MN}$, where $e_f$ is the electromagnetic charge of $q_f$. Furthermore, we shall be working in an ansatz where the axial field $(L_{M} - R_{M})/\sqrt{2}$ is zero,
which is consistent with the equations of motion. We are left with only the vector field $V_{M} = (L_{M} + R_{M})/\sqrt{2}$.
In this case the covariant derivatives become ordinary. Let us also split the scalar and tensor fields into real and imaginary parts: $X=\dfrac{X_+ + iX_-}{2};\ B_{MN} =
\dfrac{(B_+ +i B_-)_{MN}}{\sqrt{2}}$. These new fields are dual to the following operators in QCD:
\begin{align}
&\bar{q} q \leftrightarrow X_+\,,&  &\frac{1}{\sqrt{2}}\,\bar{q}\sigma_{\mu\nu} q \leftrightarrow B_{+\mu\nu}\,,\nonumber\\
&i\bar{q}\gamma_5 q \leftrightarrow X_-\,,&   &\frac{i}{\sqrt{2}}\,\bar{q}\gamma_5\sigma_{\mu\nu} q\leftrightarrow B_{-\mu\nu}\,,\nonumber\\
&&\bar{q}\gm q \leftrightarrow V_{\mu}\,.~~~~~~~~~~~~~&&\label{correspondence}
\end{align}
We have noted that $B_{\mu\nu}$ is bifundamental with respect to $U(N_f)_L\times U(N_f)_R$ which guarantees its being complex-valued, its real and imaginary parts corresponding to the tensor and pseudotensor operators (\ref{correspondence}). These operators happen to be related to each other in 4D, $\bar{q}\sigma^{\mu\nu}\gamma_5q = \dfrac{i}{2} \epsilon^{\mu\nu}_{~~\lambda\rho}\bar{q}\sigma^{\lambda\rho}q$, that fact is reflected in  Eq.\,(\ref{bmunu}). From the holographic point of view, this condition is ensured by the fact that the kinetic term for $B_{\mu\nu}$ (\ref{fullaction}) is of the first order in derivatives, which leads to its complex self-duality \cite{harvey, Karch:2011}. Thus, as we shall see, the ``double counting" of the degrees of freedom that arises after we have introduced a complex tensor field is compensated by constraints imposed on half of them, see Eqs.\,(\ref{B03eq}). One may wonder whether it is possible to avoid this redundancy by dealing with a real-valued tensor field from the beginning; however, amending the model in this way while preserving holographic field-operator correspondence rules and general self-consistency appears to be quite cumbersome.

Let us now rewrite the action (\ref{fullaction}) in terms of those fields:
\ba
{\cal S} &=& \int\ d^5x \sqrt{-g}\ {\rm Tr}\,\Big\{ -\frac{1}{4g_5^2} F_V^2 + \frac{g_B}{3}\epsilon^{MNPQR}\big( B_{-MN}H_{+PQR} - B_{+MN}H_{-PQR}\big) \nonumber\\[1mm]
 &+& 
\sum\limits_{+,-} \Big[\!- g_B m_B B_{\pm MN}B^{\pm MN}\! + \frac{g_X^2}{4} \lb \partial_M X_{\pm}\partial^M X_{\pm} -m_X^2 X_{\pm}^2 \rb
+\frac{\lambda}{2} X_{\pm}\lb F_V \rb_{MN} B_{\pm}^{MN}\Big] \Big\}.~~~\label{action}
\ea

\subsection{Equations of motion}

The action (\ref{action}) yields the following first-order equations of motion for the tensor field:
\ba
\pm z\epsilon^{\mu\nu\lambda\rho}H_{\pm z\lambda\rho} + 2 B_{\mp}^{\mu\nu} = \frac{\lambda}{4g_B} X_{\mp} F_V^{\mu\nu},\nonumber\\[1mm]
\pm \frac{z}{3}\epsilon^{\mu\lambda\rho\sigma}H_{\pm \lambda\rho\sigma} + 2 B_{\mp}^{\mu z} = \frac{\lambda}{4g_B} X_{\mp} F_V^{\mu z},\label{1stordereqs}
\ea
where the indices are contracted with a flat metric $\rm diag(+----)$. They may be rewritten as second-order equations in which the real and imaginary components are disentangled
and $(F_V)_{\mu\nu}$ is assumed to be $z$-independent (which is a self-consistent solution in the case of a constant uniform magnetic field). Along with the equation on $X$ we have:
\ba
&& z\dz\lb z H_{\pm}^{z\alpha\beta}\rb + B_{\pm}^{\alpha\beta} + z^2 \dmu H_{\pm}^{\mu\alpha\beta}
= \frac{\lambda}{8 g_B} \lbr X_{\pm} F_V^{\alpha\beta} \pm z\dz X_{\mp}\tilde{F}_V^{\alpha\beta} \rbr,\nonumber\\[2mm]
&& z^2\dl H_{\pm}^{\lambda\mu z} + B_{\pm}^{\mu z}  = \pm\frac{\lambda}{2 g_B} z\dl\lb X_{\mp} \tilde{F}_V^{\mu\lambda}\rb,\nonumber\\[2mm]
&& \dz\lb\frac{1}{z^3}\dz X_{\pm}\rb + \frac{3}{z^5} X_{\pm} - \frac{1}{z^3}\dmu\dmuu X_{\pm} = -\frac{\lambda}{g_X^2}\frac{1}{z} (F_V)_{\mu\nu}B_{\pm}^{\mu\nu}.\label{2ndordereqs}
\ea

Directing the third axis along the magnetic field so that $(F_V)_{12} = (\tilde{F}_V)_{03} = {\bf B}$ we get the following equations on $\lb B_{\pm}\rb_{12}$ and $X_{\pm}$:
\ba
&& \lb \dz^2 + \frac{1}{z}\dz - \frac{1}{z^2} - \dmu\dmuu \rb \lb B_{\pm}\rb_{12} = -\frac{\lambda}{8 g_B}\frac{1}{z^2} X_{\pm} \lb F_V\rb_{12},\label{B12eq}\nonumber\\[2mm]
&& \lb \dz^2 - \frac{3}{z} \dz + \frac{3}{z^2} - \dmu\dmuu \rb X_{\pm} = -\frac{2\lambda}{g_X^2} z^2 \lb F_V\rb_{12} \lb B_{\pm}\rb_{12}.\label{Xeq}
\ea
From Eqs. (\ref{2ndordereqs}) it follows that we will also have nontrivial $\lb B_{\mp}\rb_{03}$, $\lb B_{\mp}\rb_{0z}$, and $\lb B_{\mp}\rb_{3z}$ components,
which may be expressed through $\lb B_{\pm}\rb_{12}$ with the use of Eqs. (\ref{1stordereqs}) (assuming $\epsilon^{0123}=1$):
\be
\lb B_{\mp}\rb_{03} = \pm z\dz \lb B_{\pm}\rb_{12},\qquad \lb B_{\mp}\rb_{0z} = \pm z\partial_3 \lb B_{\pm}\rb_{12},\qquad \lb B_{\mp}\rb_{3z} = \pm z\partial_0 \lb B_{\pm}\rb_{12}. \label{B03eq}
\ee

A most general property of the equations is that the scalar and tensor degrees of freedom $X_+, B_{+12}$ decouple from the pseudoscalar and pseudotensor $X_-, B_{-12}$,
thus forming two independent sectors,
while due to complex self-duality $B_{\mp 03},\ B_{\mp 0z},\ B_{\mp 3z}$ are admixed to the first (second) sector of the solution. Those sectors may be treated independently.

\subsection{Solutions and boundary conditions}

After we Fourier-transform the equations, solutions of the Eqs. (\ref{B12eq}, \ref{Xeq}) assume the form:
\ba
X_+ + i X_- &=&  z^2 f_X(qz)e^{ikx_3-i\omega t};\nonumber\\[1mm]
\lb B_+ +i B_-\rb_{12} &=& \frac{g_X}{4\sqrt{g_B}} f_B(qz)e^{ikx_3-i\omega t},\label{solution}
\ea
where $f_X(qz)$ and $f_B(qz)$ are, generally speaking, superpositions of four Bessel functions:
\ba
f_X(qz) = C_1 {\cal J}_1\big(\sqrt{1+\beta}qz\big) + C_2 {\cal J}_1\big(\sqrt{1-\beta}qz\big) +  C_3 {\cal Y}_1\big(\sqrt{1+\beta}qz\big) + C_4 {\cal Y}_1\big(\sqrt{1-\beta}qz\big);~~~\nonumber\\[2mm]
f_B(qz) = C_1 {\cal J}_1\big(\sqrt{1+\beta}qz\big) - C_2 {\cal J}_1\big(\sqrt{1-\beta}qz\big) +  C_3 {\cal Y}_1\big(\sqrt{1+\beta}qz\big) - C_4 {\cal Y}_1\big(\sqrt{1-\beta}qz\big),~~~\label{gensolution}
\ea
where $\beta = \dfrac{|\lambda|}{2g_X\sqrt{g_B}}\lv {\bf B}/q^2\rv$ and $q^2 = \omega^2 - k^2$ is the Minkowski 4-momentum squared. ${\cal J},\ {\cal Y}$ are the Bessel and Neumann functions or their analytical continuations
(if we want to consider greater magnetic fields or solutions with Euclidean momenta). The Neumann functions in (\ref{gensolution}) correspond to non-normalizable modes of the $AdS_5$ fields (\ref{solution}) in the UV. As we are only interested in the mixing between
vacuum expectation values without sources, $C_3=C_4=0$. $C_1$ and $C_2$ are determined by the boundary conditions in the IR.

According to (\ref{B03eq}), there is a constraint that relates $\lb B_{\mp}\rb_{03}$ to $\lb B_{\pm}\rb_{12}$, which means that in order to construct a self-consistent
variation principle for the tensor field one needs to take into account that half of the tensor degrees of freedom are not independent due to the tensor field's complex
self-duality. Such variation principle has been proposed in \cite{Karch:2011}, and it states that in our case
\be
\delta_B {\cal S} = 2g_B \int d^4x \lb B_{+12} + B_{-03} \rb \delta \lb B_{12} - B_{-03} \rb .\label{variationB}
\ee
Note that according to (\ref{B03eq}) $B_{+12}$ and $B_{-03}$ have equal normalizable modes, thus contributing equally to the tensor condensate (\ref{variationB}).
Since the kinetic term of the tensor field is of the first order in derivatives, the boundary variation term in (\ref{variationB}) contains no differentiation with respect to $z$. Hence
it makes sense to impose on it a Dirichlet boundary condition at $z=z_m$ rather than a Neumann one. From (\ref{variationB}) it also follows that a Dirichlet condition has to be
imposed on the sum $\lb B_{+12} + B_{-03} \rb$. Hence,
\be 
\frac{C_1}{C_2} = \frac{{\cal J}_1\big(\sqrt{1-\beta}\,qz_m\big)+\sqrt{1-\beta}\,qz_m {\cal J}'_1\big(\sqrt{1-\beta}\,qz_m\big)}{{\cal J}_1\big(\sqrt{1+\beta}\,qz_m\big)+\sqrt{1+\beta}\,qz_m {\cal J}'_1\big(\sqrt{1+\beta}qz_m\big)} \,.
\ee

There is no infrared boundary condition for $X$, so the overall value of $C_i$ remains undetermined. Nevertheless, we can obtain the ratio of the tensor and scalar condensates
(the scalar one is determined from  (\ref{Xmatching}), while the tensor condensate is read off of the variation of the action with respect to the tensor field (\ref{variationB})):
\be
\ltr \bar{q}\sigma_{12}q \rtr \propto 8g_B\frac{g_X}{4\sqrt{g_B}} \lim\limits_{z\rightarrow 0}\frac{f_B(qz)}{z}\,;\qquad
\ltr \bar{q}q \rtr \propto g_X^2 \lim\limits_{z\rightarrow 0}\frac{f_X(qz)}{z}\,,\nonumber
\ee
hence
\be \mu({\bf B};q) = \frac{\ltr \bar{q}\sigma_{12}q \rtr}{\ltr \bar{q} q \rtr} = \frac{2\sqrt{g_B}}{g_X} \lim\limits_{z\rightarrow 0}\frac{f_B(qz;{\bf B})}{f_X(qz;{\bf B})}\,.\label{mu}\ee

\begin{figure}[h]
\includegraphics[width=\linewidth]{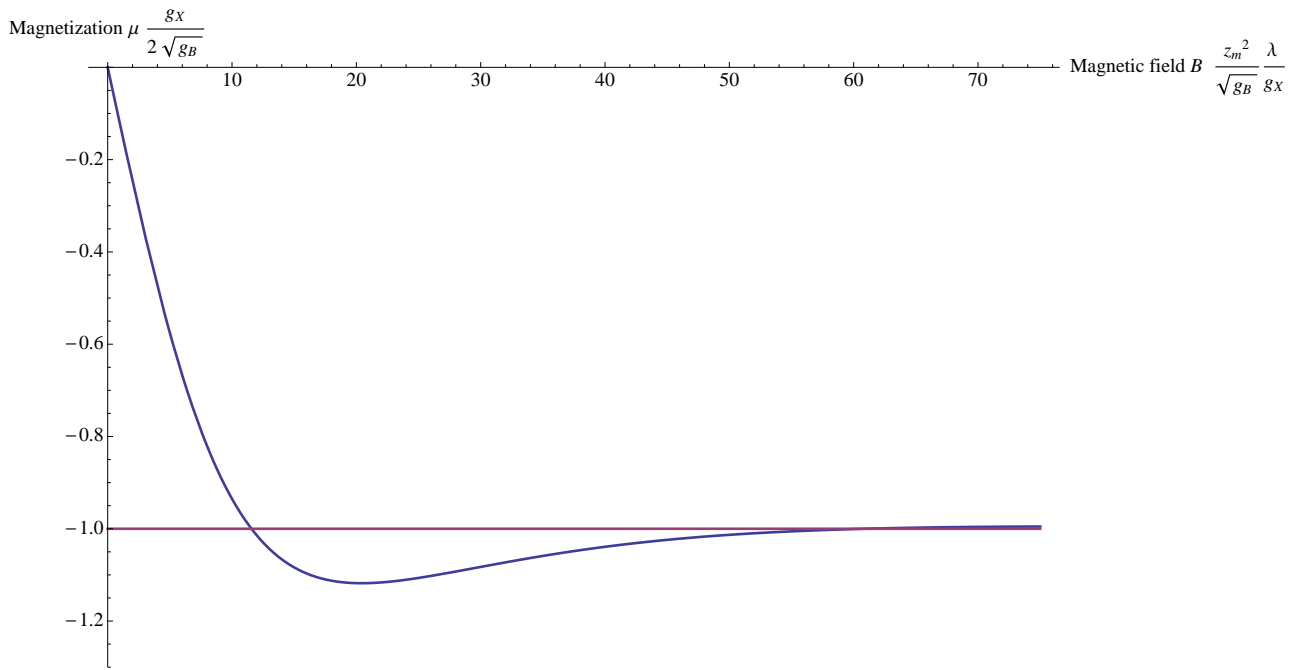}
\caption{Magnetization $\mu({\bf B})$ (blue) vs its strong
field asymptotics (red).} \label{fig_mag}
\end{figure}

\begin{figure}[h]
\includegraphics[width=\linewidth]{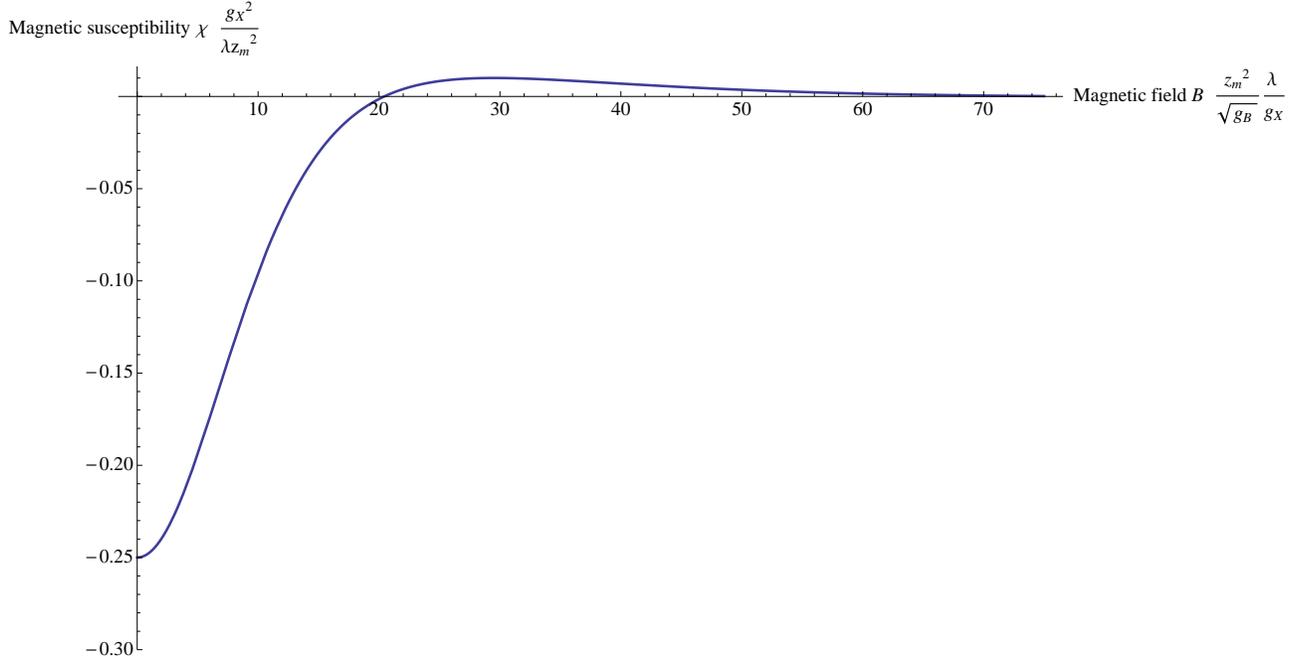}
\caption{Magnetic susceptibility of the quark condensate $\chi({\bf B})$} \label{fig_susc}
\end{figure}

Setting the 4-momentum to zero we are able to obtain the magnetization $\mu({\bf B})$ and the magnetic susceptibility $\chi({\bf B})=\dfrac{d}{d{\bf B}}\,\mu({\bf B})$ for a uniform condensate
in terms of Bessel functions of $\sqrt{\slantfrac{|\lambda|}{2g_X\sqrt{g_B}}\,{\bf B}z_m^2}$. They are presented here on Figs.\,\ref{fig_mag} and \ref{fig_susc}, respectively.

Their main properties are that 
\be 
\lim\limits_{{\bf B}\rightarrow \infty}\mu({\bf B}) = -\frac{2\sqrt{g_B}}{g_X} \ee
and 
\be 
\chi({\bf B}) = -\frac{|\lambda|}{g_X^2}\frac{z_m^2}{4}\lb 1-\frac{1}{96}\frac{\lambda^2}{g_X^2 g_B}{\bf B}^2 z_m^4 + {\cal O}\lb {\bf B}^4 z_m^8 \rb \rb, \qquad {\bf B} \rightarrow 0 . 
\ee
Note that the magnetization changes its behavior from one linear in the magnetic field to a constant at values of the magnetic field of order of ${\bf B} \sim z_m^{-2} \sim \Lambda_{QCD}^2$. Its
constant asymptotic is a behavior to be expected. At large magnetic fields the dynamics of the theory become effectively two-dimensional and the tensor chiral condensate is kinematically reduced
to a scalar one.

If we substitute the values of the constants of the model from \cite{Karch:2011} we obtain\footnote{Our constants $g_B$ and $\lambda$ are 3 times larger than those in \cite{Karch:2011}
due to the fact that our variation of the action (\ref{variationB}) with respect to the tensor field is, in a similar way, 3 times larger than the variation in \cite{Karch:2011}.}
\be
\mu(\infty) = 1/\sqrt{3}\,,\qquad \chi(0) = -\frac{z_m^2}{72}\,. \ee
Let us recall that $z_m$ is fixed by the mass of the $\rho$-meson \cite{Erlich_Katz:2005}, $z_m \sim 2.4\ m_{\rho}^{-1}$, which means
that $\chi(0) \sim -0.08\ m_{\rho}^{-2} \sim -0.13\ \rm GeV^{-2}$.
One may compare that to the results \cite{Ioffe:book}, where the susceptibility has been analyzed from the point of view of sum rules and has been determined to be
$\chi \sim -3.15 \pm 0.30\ \rm GeV^{-2}$, while the pion dominance and OPE for the $\langle VVA \rangle$ diagram give \cite{vai} $\chi \sim -8.9\ \rm GeV^{-2}$.
(Other results include a holographic computation of the $\langle VVA \rangle$ diagram, which led to a value $\chi \sim -11.5\ \rm GeV^{-2}$ \cite{gk}; the use of vector dominance \cite{rhodom}
gives $\chi \sim -(3.38 \div 5.67)\ \rm GeV^{-2}$.)  One can see that there is a large discrepancy between the numerical results that clearly requires more investigation.
However and more importantly, our results reproduce the general properties both of the susceptibility and of the magnetization -- the weak-field expansion of the former and the negative constant
asymptotic of the latter.

As for the particular  value of the magnetization, the lattice calculation yields a  different saturation value in a large field  \cite{bui}:
$\lim\limits_{{\bf B}\rightarrow \infty}\mu({\bf B}) = -1$. It
has also been discussed in the  NJL model \cite{ita}. In theory, such a value would tell us that both condensates get contributions only from the LLL (lowest Landau level)
in the strong magnetic field.
On the other hand a different saturation value obtained in this paper implies that the picture is more
complicated. Indeed, as it has been shown by Miransky {\it et al.} \cite{summation}, in  some problems the summation of the infinite number of higher Landau levels is needed to reproduce the correct result. It has also been argued in \cite{summation} that the LLL approximation is reliable only in the kinematic region when the momenta satisfy the condition $q_\perp^2 \gg q_\parallel^2$. In our particular case the problem is completely static ($q_i =0$), so this condition is not fulfilled and the careful treatment of the higher Landau levels is desirable. Another simple argument concerns the derivation of the magnetization via
the Dirac operator spectrum \cite{gk,bui}. An analogue of the Casher-Banks formula
implies that the result obtained in \cite{bui} is based on the factorization
of the product of two operators under the averaging over the gluon configuration.
The lack of factorization  could
be the origin of disagreement with the lattice result.
Anyway, this point needs further clarification.

\subsection{Vector current}

Let us now consider perturbations of the vector field about the solution $F_{V12} = {\bf B}$. They obey the following equations:
\ba
\dz\lb \frac{1}{z}\dz V_3 \rb - \frac{1}{z}\dt^2 V_3 +\frac{1}{z} \dt\dzed V_0 =
 4g_5^2 \lambda \sum\limits_{+,-} \lbr -\dz\lb \frac{1}{z}X_{\pm}B_{\pm 3z} \rb - \frac{1}{z}\dt\lb X_{\pm}B_{\pm 03} \rb \rbr\label{V3eq},\nonumber\\[2mm]
-\dz\lb \frac{1}{z}\dz V_0 \rb - \frac{1}{z}\dzed^2 V_0 +\frac{1}{z} \dt\dzed V_3 =
4g_5^2 \lambda \sum\limits_{+,-} \lbr -\dz\lb \frac{1}{z}X_{\pm}B_{\pm 0z} \rb + \frac{1}{z}\dt\lb X_{\pm}B_{\pm 03} \rb \rbr\label{V0eq}.~
\ea
One may note that the longitudinal components $B_{\pm z\mu}$ (as well as $B_{\pm 03}$) become sources for the vector current and charge density. However, the r.h.s. of the
Eqs. (\ref{V3eq},\ref{V0eq}) contains products of fields from different sectors of the solution. Furthermore, if we consider small fluctuations, the vector field turns out to be
a fluctuation of the second order. There are obvious similarities with Eq.\,(\ref{current_chipt}), where $\pi =  \arg{(X_+ + iX_-)}$.

\section{Discussion}

In this paper we have discussed a few issues concerning the magnetic susceptibility
of the quark condensate. We have shown that the SY relation which yields the value of
the susceptibility at large $Q^2$ is consistent with the chiral log counting at small $Q^2$.
The nonvanishing value of the susceptibility implies a specific term in the effective
lagrangian and we have analyzed the role of this term in the holographic approach. It turns
out that it provides  the magnetization at any value of the magnetic field. Surprisingly
the saturation value in the strong magnetic field is small and  disagrees with the lattice
simulation. This disagreement deserves additional studies.

A satisfactory explanation of the SY relation is still absent.
It implies
a peculiar relation between the kinetic and topological terms. Such relation
is natural from the brane viewpoint and one could expect a kind
of Ward identity to stand behind it. We have not found the symmetry which
would provide such a Ward identity, however more efforts could be
made in this direction and we plan to return to this point elsewhere.

Which vacuum excitations are responsible  for the magnetic
susceptibility? This
question can be rephrased as one concerning the localization of the
quarks involved into the composite operator
on some vacuum defects excited by the external
magnetic field.
The answer potentially depends
on the interpretation of the background $C_{\mu \nu}$ field.
There could be several interpretations. If it is the
two-form field in NS or RR sectors it would mean that
the F1 or D1 degrees of freedom are under the carpet. The
variant with NS $B_{\mu \nu}$ field has some trouble since
in this case we are dealing with noncommutative field theory
and the field enters other terms in the Lagrangian. Hence it is unclear
if it would be possible to separate the desired term in the effective action
in a clear-cut way. If we choose the RR two-form field $C_2$ the
product $C_2\wedge F$ follows from the CS term
immediately, however
the proportionality to the chiral condensate needs an explanation.

Finally let us comment on the possible interpretation of the tensor source as
the curvature of the graviphoton one-form RR field $C_1$: $C=dC_1$.
The degrees of freedom naturally charged with respect to the
graviphoton field are the D0 branes, hence one could question how
D0 particles or instantons are captured by the magnetic field.
The potential object which could be relevant is the dyonic
instanton, that is, a  blown up instanton with a string attached to it.
Upon the blow up it behaves as a magnetic dipole with a topological
charge.

\section{Acknowledgments}
We would like to thank P.\,Buividovich and M.\,Shifman for the useful discussions. A.G. thanks
FTPI at University of Minnesota where the part of the work has been done for the
hospitality and support. The work of A.G. is partially supported by grant CRDF-RUP2-2961-MO-09.
The research of A.K. is supported by the Dynasty Foundation, by the Ministry of Education and Science of the Russian Federation under contract 14.740.11.0347
and by the RFBR grant no. 10-02-0148. The research of P.N.K. is supported by the Dynasty Foundation and by the Ministry of Education and Science of the Russian Federation under contract 14.740.11.0081.  The work of A.V. is supported in part by the DOE grant DE-FG02-94ER40823.

\end{document}